\documentclass[12pt]{article}
\usepackage[a4paper]{geometry}
\usepackage{graphicx}
\begin{document}
\title{
\Large\bf Universality classes of three-dimensional $mn$-vector
model}
\author{M. Dudka$^{1,2}$, Yu. Holovatch$^{1,2,3}$, and T Yavors'kii $^{3,4}$%\email[]{maxdudka@icmp.lviv.ua}
\\[3mm] \normalsize $^{1}$ Institute for Condensed Matter Physics, %\\ \normalsize
 Nat.
Acad. of Sci. Ukraine, \\ \normalsize UA--79011 Lviv, Ukraine\\
\normalsize $^{2}$ Institute f\"ur Theoretische Physik,
%\\ \normalsize
 Johannes Kepler Universit\"at Linz, \\ \normalsize A--4040 Linz,
Austria
 \\  \normalsize $^{3}$ Ivan Franko National University of Lviv, \\ \normalsize UA--79005
Lviv, Ukraine\\  \normalsize $^4$ Department of Physics,
University of Waterloo, \\ \normalsize N2L 3G1 Waterloo, Ontario,
Canada \\  \small Emails: maxdudka@icmp.lviv.ua, hol@icmp.lviv.ua,
tarasyk@ktf.franko.lviv.ua}
\date{\small(\today)}
\maketitle
\begin{abstract}
{ We study the conditions under which the critical behavior of the
three-dimensional $mn$-vector model does not belong to the
spherically symmetrical universality class. In the calculations we
rely on the field-theoretical renormalization group approach in
different regularization schemes adjusted by resummation and
extended analysis of the series for renormalization-group
functions which are known for the model in high orders of
perturbation theory. The phase diagram of the three-dimensional
$mn$-vector model is built marking out domains in the $mn$-plane
where the model belongs to a given universality class.\\[5mm] PACS
numbers:  {05.50.+q, 64.60.Ak}}
\end{abstract}
%\pacs{05.50.+q, 64.60.Ak}

According to the universality hypothesis \cite{univ}, asymptotic
properties of the critical behavior remain unchanged for different
physical systems if these are described by the same global
parameters. The field-theoretical renormalization group (RG)
approach \cite{RGbooks} naturally takes into account the global
parameters and derives properties of critical behavior from long
distance properties of effective field theories.  In the present
paper we study the long-distance properties of the
$d=3$-dimensional $mn$-vector model which is introduced by the
following effective field-theoretical Hamiltonian
 \cite{Guillou74}:
\begin{eqnarray} \label{Heff}
\!{\cal H}[\phi(x)]\!\!\!&{=}&\!\!\!\!\int\!{\rm d}^d x \Big\{
{1\over 2} \sum_{\alpha{=}1}^{n}\! \left[|\nabla
\vec{\phi}^\alpha|^2{+}
\mu_0^2 |\vec{\phi}^\alpha|^2\right]\!{+} %\nonumber\\&&
{u_{0}\over 4!} \sum_{\alpha{=}1}^{n}\!\left(|\vec{\phi}^\alpha|^2
\right)^2\! {+}
{v_{0}\over 4!} \left(\sum_{\alpha{=}1}^{n}|\vec{\phi}^\alpha|^2
\right)^2\! \Big\}\, .
\end{eqnarray}
Here,
$\vec{\phi}^{\alpha}=(\phi^{\alpha,1},\phi^{\alpha,2},\dots,\phi^{\alpha,m})$
is a tensor field of the dimension $n$ and $m$ along the first and
the second indices; $u_{0}$ and
$v_{0}$ are bare couplings; $\mu_0^2$ is a bare mass squared
measuring the temperature distance to the critical point.

Depending on the choice for the parameters $m$ and $n$, the
$mn$-vector model (\ref{Heff}) is known to describe phase
transitions of various microscopic nature. The choice $n=1$
comprises a bunch of systems that are characterized by an
$O(m)$-symmetric order parameter, while the limiting cases
$n\rightarrow 0$ and $n\rightarrow \infty$ correspond to these
systems exposed to the quenched  \cite{Grinstein76} and annealed
 \cite{Emery75} disorder respectively. The choice $m=1$, arbitrary
$n$ corresponds to the cubic model  \cite{Aharony76}. A separate
interest is provided by the case $m=2,\, n=2$ describing
 \cite{Mukamel75} helical magnets  and antiferromagnetic phase
transitions in ${\rm TbAu_2}$, ${\rm DyC_2}$ as well as by the
case $m=2,\, n=3$ describing antiferromagnetic phase transitions
in ${\rm TbD_2}$, ${\rm Nd}$.

All of the mentioned cases of the $mn$-vector model were subjects
of separate extensive studies (see e.~g. Refs.
 \cite{Pelissetto02,PRB_cub,our_review} and references therein).
They led to a consistent description of criticality in the $O(m)$
and cubic systems. In particular, the precise estimates of the
critical exponents of the cubic and of the random Ising model were
established both within high-order expansions of the massive and
minimal subtraction field-theoretical RG schemes
 \cite{Pelissetto02,our_review}. On the contrary, the cases $m=2,\,
n=2,3$ remain to be controversial. Using general non-perturbative
considerations it was shown  \cite{Cowley78} that the theory
(\ref{Heff}) belongs to the $O(2)$ universality class. On the
other hand the perturbative field-theoretical RG approach yielded
mixed data, neither proving nor rejecting this result
 \cite{Pelissetto02,mn}.

The studies infer that an intrinsic feature of the theory
(\ref{Heff}) is an interplay between the $O(k)$ (``trivial'')
universality class (with $k$ being dimensions $m$,  or $mn$) and a
new universality class. In this paper we address two problems that
concern the crossovers in the $mn$-vector model and still attract
attention. Firstly, we aim to obtain a map of universality classes
of the theory (\ref{Heff}) in the whole plane $m\geq 0,\,n\geq 0$.
Such an analysis has been performed so far in the one-loop
approximation  \cite{Aharony76}. We base the analysis on the
high-loop expansions for the RG functions of the model
(\ref{Heff}) and its special cases; in order to refine the
analysis we exploit Pad\'e-Borel resummation
\cite{Baker81,Baker78}
 of the (asymptotic)
series under consideration. Secondly, we focus attention on the
case $m=2$, $n=2,\,3$ in order to explain why the highest orders
of perturbation theory have not allowed so far to resolve what
universality class is realized in the theory. We perform analysis
in different perturbative schemes and show that only certain of
them give reliable answer.

We analyze the theory (\ref{Heff}) applying the field-theoretical
RG approach \cite{RGbooks} within weak coupling expansion
techniques. In the approach, a critical point corresponds to a
reachable and stable fixed point (FP) of the RG transformation of
a field theory. A FP $\{u^*, v^*\}$ is determined as a
simultaneous zero of the $\beta$-functions describing the change
of the renormalized couplings $u$ and $v$ under RG transformations
and being calculated as perturbative series in renormalized
couplings. The equations for the FP read:
\begin{equation}\label{FP} \left\{ \begin{array}{ccccc}
\beta_u(u^*,v^*) &=& u^*\, \varphi(u^*,v^*) &=& 0\,,
\vspace{4mm}
\\
\beta_v(u^*,v^*) &=& v^*\, \psi(u^*,v^*) &=& 0\,,
\end{array}
\right.
\end{equation}
where we have explicitly shown that the structure of the
$\beta$-functions allows their factorization for the effective
Hamiltonian (\ref{Heff}). We make use of both the dimensional
regularization with minimal subtraction  \cite{tHooft72} and the
fixed dimension renormalization at zero external momenta and
non-zero mass (massive)  \cite{Parisi80} schemes. More precisely,
we rely on the expansions for the $\beta$-functions that are known
at $d=3$ with the accuracy of six loops in the massive scheme
 \cite{Pelissetto_mn} and with five loop accuracy for the cases of
$O(m)$-vector and cubic models in the minimal subtraction scheme
(Refs.  \cite{Kleinert91} and  \cite{Kleinert95} correspondingly).

Technically, the Eqs. (\ref{FP}) for the FP can be solved in two
complementary ways. A perturbative solution is obtained by an
expansion of the FP coordinates in a small parameter
($\varepsilon=4-d$, with $d$ being the space dimension of the
model  \cite{Wilson}, in the minimal subtraction or massive
schemes, or an auxiliary pseudo-$\varepsilon$ parameter
 \cite{pseps} in the massive scheme) around the Gaussian solution
$\{u^* =0, v^* = 0\}$. Such a way formally guarantees that the
structure of solutions for the FPs remains the same after account
of higher-order contributions once it has been established in the
one-loop approximation. An alternative method (the 3d approach)
consists in the solution of Eqs.~(\ref{FP}) numerically
 \cite{Parisi80,Schloms} at a given order of perturbation theory
and provides less control on a loopwise upgrade.

Within the perturbative approach, the conditions on $m$ and $n$
under which the critical behavior of the $mn$-vector model
(\ref{Heff}) belongs to a non-trivial universality class are known
as Aharony conjecture and read  \cite{Aharony76}:
\begin{equation}
\label{cond} n_c<mn<m_cn, \qquad n>1.
\end{equation}
Here, $n_c$ and $m_c$ stand for the marginal dimensions of the
cubic model and of the random $m$-vector model. The conjecture is
based on the one-loop stability analysis of four FP solutions
compatible with the Eqs. (\ref{FP}). At $d<4$, these are the
Gaussian FP {\bf G} $\{ u^*=0, v^*=0 \}$, the FPs {\bf
P$\rm_{O(mn)}$} $\{u^*=0, v^* \neq 0\}$ and {\bf  P$\rm _{O(m)}$}
$\{u^*\neq 0, v^*=0 \}$ describing theories with one $\phi^4$
coupling and thus corresponding to the $O(mn)$ and $O(m)$
universality classes, and, finally, the mixed FP {\bf M}
$\{u^*\neq 0, v^*\neq 0\}$. It is the stability of the FP {\bf M}
that is necessary for the appearance of a new non-trivial critical
behavior.

The 3d analysis of the theory (\ref{Heff}) is obscured by our
observation that at some choice of $m$ and $n$ more than four
solutions for the FP are obtained. To convince ourselves that some
of them are not a by-product of application of resummation
procedures we propose to use the following argumentation.
According to the basics of the RG theory, at the upper critical
dimension $d=4$ any $\phi^4$ theory is governed by the Gaussian FP
 \cite{RGbooks}. Therefore, any non-Gaussian solution at $d=4$ is
out of physical interest. If such a solution survives at any $d<4$
and particularly at $d=3$, we find natural to consider it
physically meaningless by continuity. The situation becomes less
clear if a FP cannot be continually traced back to a certain
solution at $d=4$ because it disappears at some $3< d_c < 4$. In
this case the stability of the estimate for $d_c$ as well as of
the FP coordinates against application of different resummation
procedures in different orders of perturbation theory might serve
the purpose. It is to note here, that the special case of the
theory (\ref{Heff}) with $m=2,\,n=2$ is known to have exact
mapping onto the model describing non-collinear magnetic ordering
 \cite{Pelissetto02}. Within the massive RG scheme, standard
six-loop 3d analysis of this model allowed to find a stable FP
which does not have the counterpart within the perturbative
$\varepsilon$-expansion  \cite{Pelissetto01}. But one can not
follow the evolution of the FP as $d$ approaches 4 because in this
case the resummation procedure is ill-defined
\cite{Pelissetto03}.

We use both perturbative and 3d analysis as complementary ways to
establish the map of universality classes of the theory
(\ref{Heff}). We find that, in addition to the conditions
(\ref{cond}), the {\it high}-order map is controlled by a
degeneracy condition of {\it one}-loop equations for the FP
 \cite{Dudka02}:
\begin{equation}\label{deg}
n=\frac{16(m-1)}{m(m+8)}\,.
\end{equation}
Unlike order-dependent estimates for the marginal dimensions $m_c$ and $n_c$,
this equation is independent of the order of perturbation theory and is exact.
We also observe that the
results obtained with the account of high-order contributions
differ qualitatively from those obtained in the one-loop
approximation. We consider worth to mention three peculiarities.

(i) We find a domain in the $mn$-plane where the high-loop
resummed $\beta$-functions produce no solution for the FP while
such a solution exists in the one-loop approximation. In the
$mn$-plane the domain spans from the vicinity of the point $\{
m=m_c, n=n_c/m_c \}$ upwards. There, we can solve the Eqs.
(\ref{FP}) for the mixed FP reliably neither numerically at the
fixed space dimension $d=3$ nor by application of the
pseudo-$\varepsilon$ expansion. In particular, though the
pseudo-$\varepsilon$ expansion can be formally constructed there,
its analysis by means of Pad\'e \cite{Baker81} or
Pad\'e-Borel-Leroy \cite{Baker78} technique produces highly
chaotic values both for mixed FP coordinates and its stability
exponents. (ii) We find a domain in the $mn$-plane where the 3d
analysis reveals two solutions for the mixed FP {\bf M}
co-existing in opposite quadrants of the $uv$-plane. In the
$mn$-plane the domain is located below the point $\{ m=m_c,
n=n_c/m_c \}$. Yet, we are always able to establish that one of
the two solutions is unphysical in the sense explained above. The
described phenomenon is quite stable with respect to the order of
perturbation theory and to the type of the resummation procedure
applied. In the perturbative approach only one solution for the
mixed FP is present. (iii) We observe that a smooth change of
parameters $m,\,n$ in the $mn$-plane can show up as a complex
abrupt trajectory of the FP {\bf M} in the $uv$-plane.

\begin{figure*}[th]
\begin{picture}(350,200)
\put(0,15){\includegraphics[width=60mm]{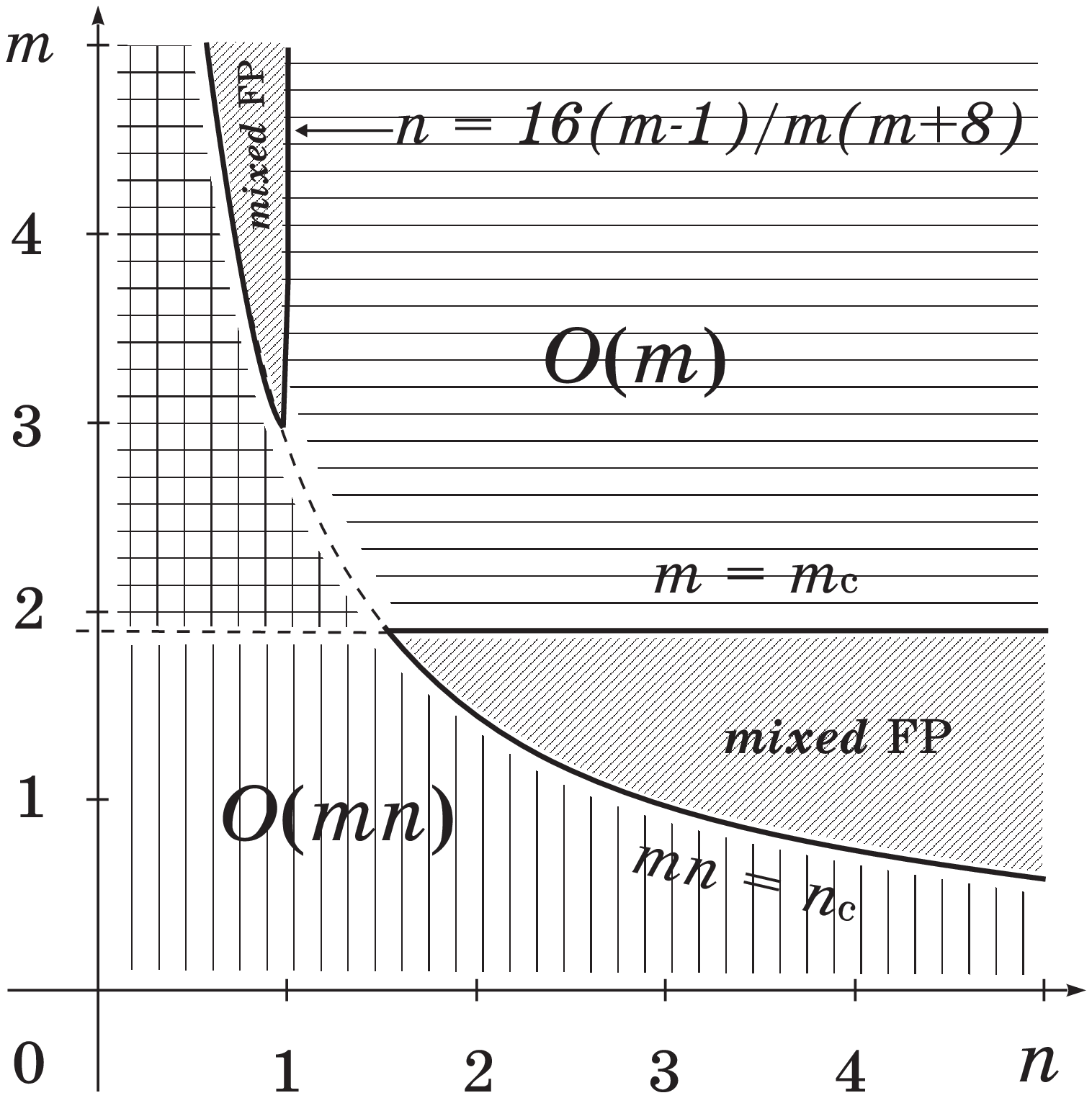}}
\put(208,15){\includegraphics[width=60mm]{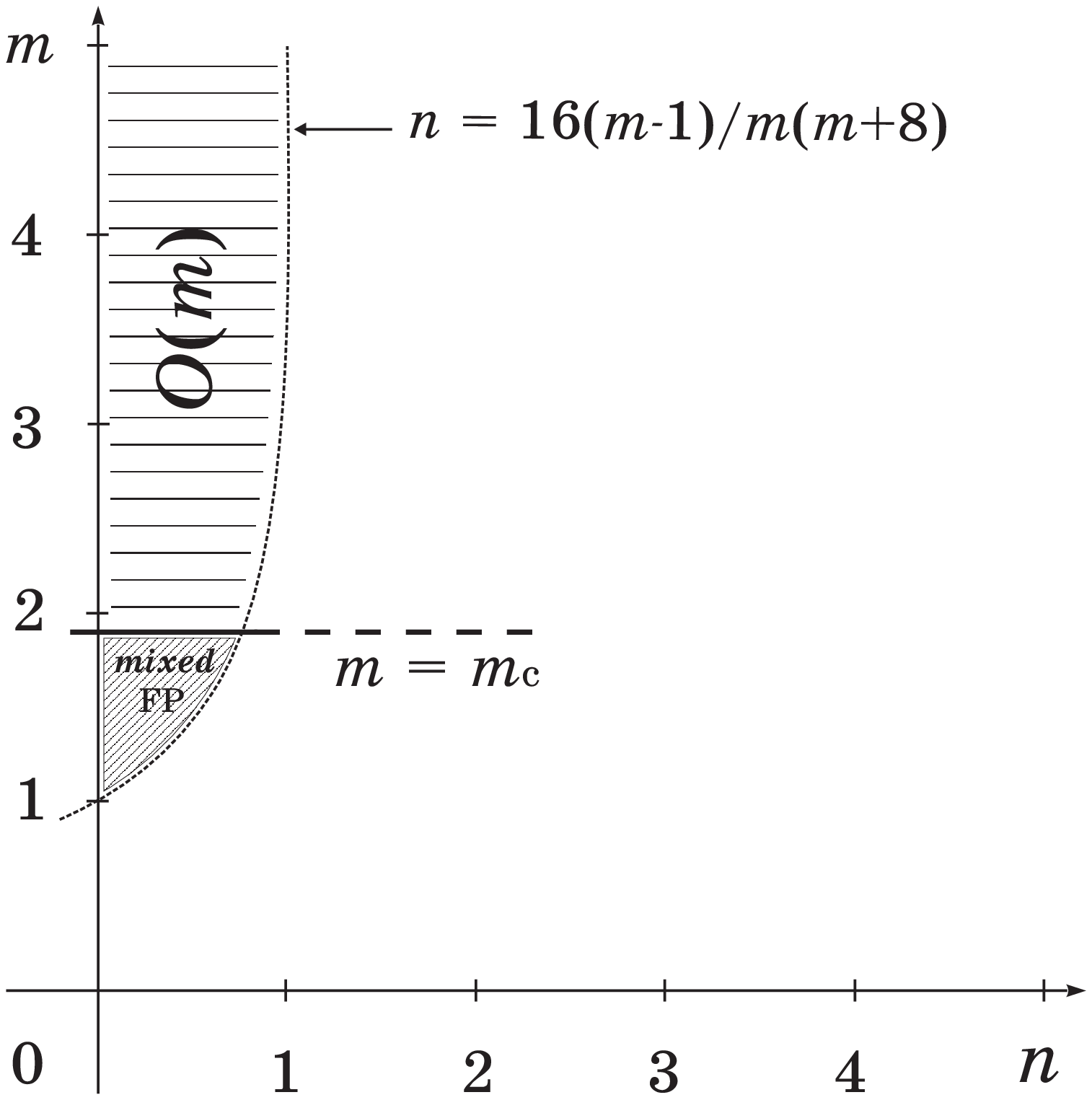}}
\put(87,0){a} \put(295,0){b}
\end{picture}
\caption{\label{mn.eps} The domains of FPs stability for the
$mn$-vector model with different signs of couplings: $\forall\,
u,\,\,v\geq 0$ (Fig. a) and  $u>0,\,\,v\leq 0$ (Fig. b). The mixed
FP is stable for values $m$ and $n$ from dark regions. The
stability regions of $O(m)$ and $O(mn)$ FPs are shaded by
horizontal and vertical lines respectively.}
\end{figure*}

Realization of various universality classes of the theory
(\ref{Heff}) besides universal Eqs.~(\ref{cond})-(\ref{deg})
depends on non-universal initial conditions for couplings. Certain
physical interpretations of the $mn$-vector model (\ref{Heff})
impose restrictions for the signs of the couplings. Namely, a
group including the cubic model ($m=1, \forall\, n$) and the cases
$m=2,\,n=2,3$ imply $u_0$ of any sign and $v_0> 0$
 \cite{Aharony76,Mukamel75}, whereas the microscopic base of the
weakly diluted quenched $m$-vector model strictly defines
$u_0>0,\,v_0\leq 0$  \cite{Grinstein76}. Taking into account such
a division along with the pseudo-$\varepsilon$ expansion based
estimates \cite{note}  for the marginal dimensions
$n_c=2.862\pm0.005$  \cite{PRB_cub} and $m_c=1.912\pm0.004$
 \cite{Dudka01}, we arrive at the high-loop map of the universality
classes of the theory (\ref{Heff}) as shown in the
Fig.~\ref{mn.eps}. There, the domains governed by different
universality classes are bounded by lines for marginal dimensions
and the degeneracy line. The FP {\bf M} is stable for values $m$
and $n$ contained in dark regions. The stability regions of the
FPs {\bf  P$\rm _{O(m)}$} and {\bf P$\rm _{O(mn)}$} are
horisontally and vertically hatched. In the cross-hatched region
in the Fig.~\ref{mn.eps}a both $O(m)$ and $O(mn)$ FPs are stable.
Here, the choise of the universality class depends on the initial
values of couplings $u,\,v$. They can be located in one of the two
domains of $uv$-plane created by the separatrix, which is
determined by the unstable mixed FP. The blank region in
Fig.~\ref{mn.eps}b denotes the region of runaway solutions. Let us
note that runaway solutions exist for the cubic-like models
(Fig.~\ref{mn.eps}a), too; however, there still exist regions of
initial couplings $u,\,v$ starting from which the stable FP is
attained.

As we mentioned above the high-loop analysis of the theory
(\ref{Heff}) encounters difficulties in some domains of the
$mn$-plane. In particular, these are the domains where the mixed
FP either disappears or can be given by two (physical and
unphysical) solutions. Our direct calculations show that such
domains (mainly) inset the regions where the FP {\bf M} is
expected to be unstable according to the
Eqs.~(\ref{cond})-(\ref{deg}) and thus does not influence the
analysis of the Fig.~\ref{mn.eps}. However, even if the solution
for the  FP {\bf M} is steadely recovered, its stability analysis
is obscured for some values of $m,\, n$. In particular, the latter
is observed for the physically interesting cases $m=2,\,n=2,\,3$.
At the rest of this paper, we aim to show that the reliability of
the stability analysis depends on the choice of a series that is
assumed as its basis.

Indeed, the stability of a FP is governed by the condition
$\Re(\omega_i)>0$ with the stability exponents $\omega_i$ being
the eigenvalues of the matrix of derivatives $B_{ij}=\partial
\beta_{u_i}/\partial u_j$ ($u_i=u,v$) taken at the FP. For the
case under consideration, \mbox{$m=2$}, \mbox{$n=2,\,3$}, one of
the eigenvalues ($\omega_2$) is large and positive both at the FPs
{\bf  P$\rm _{O(m)}$} and {\bf  M}, so it is the sign of
$\omega_1$ that controls the stability of a FP. The exponent
appears to be very small: an adjusted analysis  of the 3d six-loop
resummed RG expansions results in \cite{Pelissetto02}
 $\omega_1(m=2,\forall n)=0.007(8)$ for the FP {\bf  P$\rm _{O(m)}$}
thus providing no definitive answer about its sign. The behavior
of $\omega_1$ in different orders of perturbation theory can be
explicitly demonstrated expanding the exponent at $d=3$ in the
pseudo-$\varepsilon$ expansion \cite{pseps} parameter $\tau$ up to
the six-loop order:
\begin{eqnarray}\label{omega}
\omega_1(m=2,\forall n)&=&- 1/5 \tau{+}
0.186074\tau^2{-}0.000970\tau^3{+} 0.027858\tau^4{-}\nonumber\\ &&
0.014698\tau^5{+} 0.028096\tau^6
\end{eqnarray}
and making an attempt to evaluate the exponent at $\tau=1$  on the
base of the Pad\'e table  \cite{Baker81}: $$ \left[\! \begin
{array}{rrrrrr} {-} 0.2000&{-} 0.0139&{-}
 0.0149& 0.0130&{-} 0.0017& 0.0264
\\
{-} 0.1036&{-} 0.0149& {-}0.0140& 0.0033 & {0.0079}&o\\ {-}
0.0717& ^{0.0251}&
 ^{0.0053}& {0.0209}&o&o\\
 {-}0.0537&
 {-}0.0029& ^{0.0113}&o&o&o\\
 {-} 0.0430&
 0.0630&o&o&o&o\\
 {-}0.0351&o&o&o&o&o
\end {array}\!\right ]\!\!.
$$ In the table, numbers of the row and of the column correspond
to the orders of denominator and numerator of appropriate Pad\'e
approximant for the exponent  (\ref{omega}), the small numbers
denote unreliable data, obtained on the base of pole-containing
approximants, $o$ means that the approximant can not be
constructed. One can see, that the table shows no convergence even
along the main diagonal and those parallel to it, where the Pad\'e
analysis is known to provide the best convergence of results
 \cite{Baker81}.

On the contrary, if one first defines a pseudo-$\varepsilon$
series for the value  $m=m_c$ where the exponent
$\omega_1(m,\forall n)$ changes its sign, one gets the series
which has much better behavior  \cite{Dudka01}:
\begin{eqnarray}
\label{mc} m_c&=&4-8/3\tau+ 0.766489{\tau}^{2}-
0.293632{\tau}^{3}+%\nonumber\\&&
0.193141{\tau}^{4}-0.192714{\tau}^{5}.
\end{eqnarray}
Indeed, the corresponding Pad\'e table for $m_c$ reads:
$$ \left
[\begin {array}{cccccc} 4& {1.3333} & {2.0998}& {1.8062}& {1.9993}
& {1.8066}\\ \noalign{\medskip} 2.4& 1.9287& 1.8875& 1.9227& {
1.9029}&{\it o}\\ \noalign{\medskip} 2.0839& ^{1.8799}& 1.9084&
1.9085&{\it o}&{\it o}\\
 \noalign{\medskip} 1.9669& 1.9311& 1.9085&
 {\it o}&{\it o}&{\it o}\\
\noalign{\medskip} 1.9398& ^{2.2425}&{\it o}&
{\it o}&{\it o}&{\it o}\\
\noalign{\medskip} 1.9106&{\it o}&{\it o}&
{\it o}&{\it o}&{\it o}
\end {array}
\right ]\, $$ and leads to the conclusion $m_c<2$ already in the
three-loop order (c.f. the convergence of the results along the
diagonals of the table). A more efficient Pad\'e-Borel-Leroy
resummation procedure applied to the series (\ref{mc}) results in
an estimate \cite{Dudka01} $m_c=1.912\pm0.004$. From here one
concludes that $\omega_1(m=2,\forall n)>0$, the FP {\bf P$\rm
_{O(m)}$} at $m=2, n=2,\, 3$ is stable, and governs the critical
behavior of the $mn$-vector model. In this way the perturbative RG
scheme leads to the results which are in agreement with general
considerations of Ref.  \cite{Cowley78}, where it was shown that
the theory (\ref{Heff}) belongs to the $O(2)$ universality class
for these field dimensions.

Carrying out an analysis of conditions upon which the $mn$-vector
model belongs to the given universality class we met two problems
which are worth to be mentioned at the concluding part of this
paper. The first is that an analysis of the resummed RG functions
directly at fixed space dimensions may lead to an appearence of
the unphysical FPs. One of the ways to check the reliabily of an
analysis is to keep track of the evolution of the given FP with
continuous change of $d$ up to the upper critical dimension $d=4$.
The second observation concerns analysis of the FP stability:
taking into considerations the contradictory results obtained by a
direct analysis of the stability exponents we suggest that the
most reliable way to study the boundaries of universality classes
in field-theoretical models with several couplings consists in an
investigation of the expansions for marginal dimensions. We
believe that these our observations might be useful at the
analysis of critical propeties of other field-theoretical models
of complicated symmetry.

Work of Yu.H. and M.D. was sup\-por\-ted in part by the
Aus\-tri\-an Fonds zur F\"orderung der wissen\-schaft\-lichen
For\-schung, project No 16574-PHY and by the French-Ukrainian
cooperation Dnipro project.

\end{document}